\documentclass[apl,showpacs,twocolumn]{revtex4}
\usepackage{tipa}
\usepackage{txfonts}
\usepackage{amssymb}
\usepackage{graphicx,color,ulem}
\begin{document}


\title{Superconductivity Induced in Iron Telluride Films by Low Temperature Oxygen Incorporation}
\author{Yuefeng Nie$^{1,2}$}
\author{Donald Telesca$^{1}$}
\author{Joseph I. Budnick$^{1,2}$}
\author{Boris Sinkovic$^{1}$}
\author{Barrett O. Wells$^{1,2}$}
\email{wells@phys.uconn.edu}
\affiliation{$^{1}$Department of
Physics, University of Connecticut, Storrs, CT 06269, USA}
\affiliation{$^{2}$Institute of Materials Science, University of
Connecticut, Storrs, CT 06269-3136, USA}


\begin{abstract}
We report superconductivity induced in films of the
non-superconducting, antiferromagnetic parent material FeTe by low
temperature oxygen incorporation in a reversible manner. X-ray
absorption shows that oxygen doping changes the nominal Fe valence
state from 2+ in the non-superconducting state to mainly 3+ in the
superconducting state. Thus superconductivity in O doped FeTe occurs
in a quite different charge and strain state than the more common
FeTe$_{1-x}$Se$_x$. This work also suggests a convenient path for
conducting doping experiments in-situ with many measurement
techniques.
\end{abstract}

\pacs{74.70.Xa, 74.78.-w, 78.70.Dm}

\maketitle



The discovery of LaO$_{1-x}$F$_x$FeAs high-temperature
superconductors~\cite{Hosono01} has aroused great interest in
iron-based superconductivity. Besides the original 1111-type
$\textit{Re}$FeAsO ($\textit{Re}$ = rare earths), the 111-type
$\textit{A}$FeAs ($\textit{A}$ = alkali metal), 122-type
$\textit{Ae}$Fe$_2$As$_2$ ($\textit{Ae}$ = alkaline earths), and
11-type Fe$\textit{X}$ ($\textit{X}$ = chalcogens) are oxygen free.
The latter system is structurally the simplest, consisting only of
FeX buckled planes. For the Fe-based superconductor families, there
are two different ways in which superconductivity was achieved:
chemical or physical pressure and charge doping. The original
LaO$_{1-x}$F$_x$FeAs compound is charge doped versus the LaOFeAs
parent compound. In contrast, pressure induces superconductivity in
the undoped compounds of 1111-type LaOFeAs~\cite{Okada02} and
122-type AFe$_2$As$_2$ (A =Sr, Ba)~\cite{ParkT01,Alireza01}, but
decreases the the critical temperature of 111-type
LiFeAs~\cite{Gooch01}.

For the iron chalcogenides, FeTe is considered the parent compound
as it is antiferromagnetic but not superconducting. Chemical
pressure via isoelectronic substitution of Se or S onto Te sites
produces a superconductor~\cite{Hsu01,Fang01,Mizuguchi01}. FeSe
shows the most dramatic pressure effect among the Fe-based
superconductors with the T$_{onset}$ increases from 8.5 to 36.7 K
under 8.9 GPa~\cite{Medvedev01}. However, no form of charge doping
or of physical pressure on FeTe has been shown to produce a
superconductor. Recently, superconductivity has been reported in
FeTe films and the result attributed to the in-plane lattice
strain~\cite{HanY01}. Exactly what aspect of doping induces
superconductivity from a parent compound in the Fe-based
superconductors is unknown. In the Fe-chalcogenide system the
primary discussions have focused on the effects of distorting the
lattice~\cite{Yeh01,Hsu01,Wang01}. More broadly, there has been some
discussion that since both isovalent distortions and charge doping
can produce a superconductor, then the key feature must be avoiding
the antiferromagnetically ordered state rather than reaching any
particular place in a charge-strain phase diagram.

In this letter, we report a study of films of the parent compound
FeTe which we make superconducting by incorporating oxygen after
growth. We first observed superconductivity after long term exposure
to the ambient atmosphere at room temperature. Careful low
temperature annealing experiments confirm that superconductivity is
established by the incorporation of excess oxygen in a reversible
manner. The superconductivity we induce in FeTe films has an onset
temperature around 10 K. Spectroscopic measurements confirm that
oxygen is being taken up by the bulk of the film structure and
alters the nominal Fe valency, thus indicating a very different path
towards achieving superconductivity from the FeTe parent than
substitutionally doping Se on Te sites.


\begin{figure}
\begin{center}
\includegraphics[width=80mm]{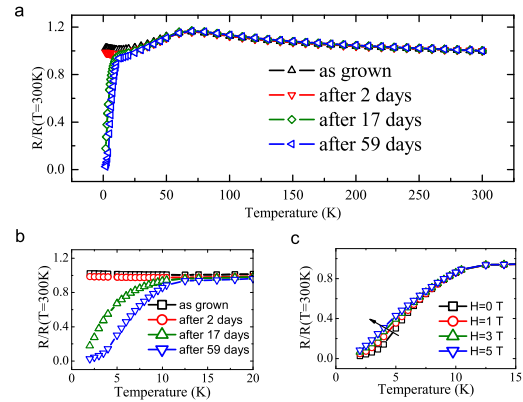}
\renewcommand{\figurename}{Figure}
\caption{ \label{fig:fig1.eps} {\bf Resistivity measurements for
FeTe films with various amounts of air exposure.} The resistivity is
normalized to the resistivity at 300 K. {\bf (a)}. A superconducting
transition appears by 17 days of air exposure. The close up of the
transition region in {\bf (b)} shows that the superconducting
transition continues to sharpen for up to two months in air reaching
essentially zero resistance at 2 K. {\bf (c)} shows the magnetic
field dependence of the resistivity vs temperature with the expected
result that the superconducting transition is suppressed in an
applied field.}
\end{center}
\end{figure}

FeTe films were grown on MgO (001) single crystal substrates by
pulsed laser deposition techniques, using a nominal 1:1 FeTe target.
The base pressure of the system is around 7x10$^{ - 8}$ torr. During
deposition, the substrate temperature was kept at 380 $^{0}$C in a
vacuum of better than 2x10$^{ - 7}$ torr. The deposition rate is
around 1.7 nm per minute. After the deposition, the samples were
cooled down to room temperature at a rate of 4 $^{0}$C/min in
vacuum.

Film thickness was measured using a laser interferometer. The
resistance of the films was measured by using the 4-point technique
with silver paint contacts made at room temperature. Temperatures
and magnetic fields were obtained using the cryostat and magnet of a
commercial magnetometer. Crystalline quality and orientation were
examined by a two-circle X-ray diffractometer with a Cu K$_\alpha$
source and a four circle diffractometer with an area detector and a
Cr K$_\alpha$ source. High resolution measurements of the lattice
constants were taken at beamline X22C of the National Synchrotron
Light Source (NSLS) at Brookhaven National Laboratory. X-ray
absorption (XAS) experiments were performed at NSLS U4B beamline.
The XAS spectra were recorded simultaneously in total electron yield
(TEY) and total florescence yield (TFY).

Figure 1 shows the effect of exposure to air on the resistivity.
Overall, the resistivity of these materials is weakly metallic,
consistent with bulk materials~\cite{Hu01,Chen01,Mizuguchi01}. In
addition, all of the resistivity vs temperature curves show a broad
peak around 70K, an effect also seen in the bulk and associated with
the concurrent antiferromagnetic and structural
transition~\cite{Hu01,Chen01,Wang01}. After two days of exposure to
air, there is almost no change observed. However, the key result for
this paper is that a superconducting transition appears by 17 days
of air exposure. The close up of the transition region in Fig. 1(b)
shows that the superconducting transition continues to sharpen for
up to two months in air reaching essentially zero resistance at 2 K.
Fig. 1(c) shows the expected magnetic field dependence that the
superconducting transition is suppressed in an applied field. The
FeTe film with varies thicknesses, range from 30 nm to 150 nm, show
essentially identical results.

\begin{figure}
\begin{center}
\includegraphics[width=80mm]{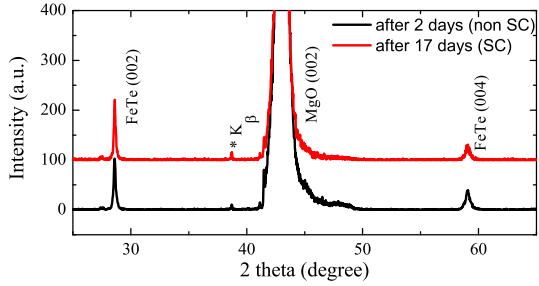}
\renewcommand{\figurename}{Figure}
\caption{\label{fig:fig2.eps} {\bf X-ray diffraction profiles of
FeTe films taken at room temperature.} The wide range profiles show
only the tetragonal (00L) peaks of FeTe films and the MgO substrate
(002) peak, indicating c-axis preferred orientation. No sign of
extra peaks of the film with longer exposure to air suggests no
structure change after incorporating oxygen from the air.}
\end{center}
\end{figure}


X-ray diffraction data confirm that the FeTe structure was always
found to be a tetragonal phase consistent with bulk measurements.
High resolution data obtained from the synchrotron light source gave
the lattice constants for the as-grown FeTe films at room
temperature as a = 3.8358(2) $\AA$, and c = 6.2734(1) $\AA$. Figure
2 shows a wide range X-ray profile in the (00L) direction for the
same FeTe film whose resistivity is shown in Fig. 1. The two data
sets are after the film was exposed to air for 2 days
(non-superconducting) and for 17 days (superconducting).  The only
peaks observable are the tetragonal FeTe (00L) peaks and the
substrate MgO (002) peak. This scan indicates the basic epitaxial
nature of the film and shows that the structure does not break down
as oxygen is incorporated and superconductivity sets in. Further,
clear shifts of FeTe (00L) peaks were observed for the films
annealed in oxygen atmosphere indicating that the c lattice constant
increases with the incorporation of oxygen. The amount of increase
in the c axis depends upon the amount of excess oxygen added,
varying from 0.01 $\AA$ to 0.03 $\AA$.

\begin{figure}
\begin{center}
\includegraphics[width=80mm]{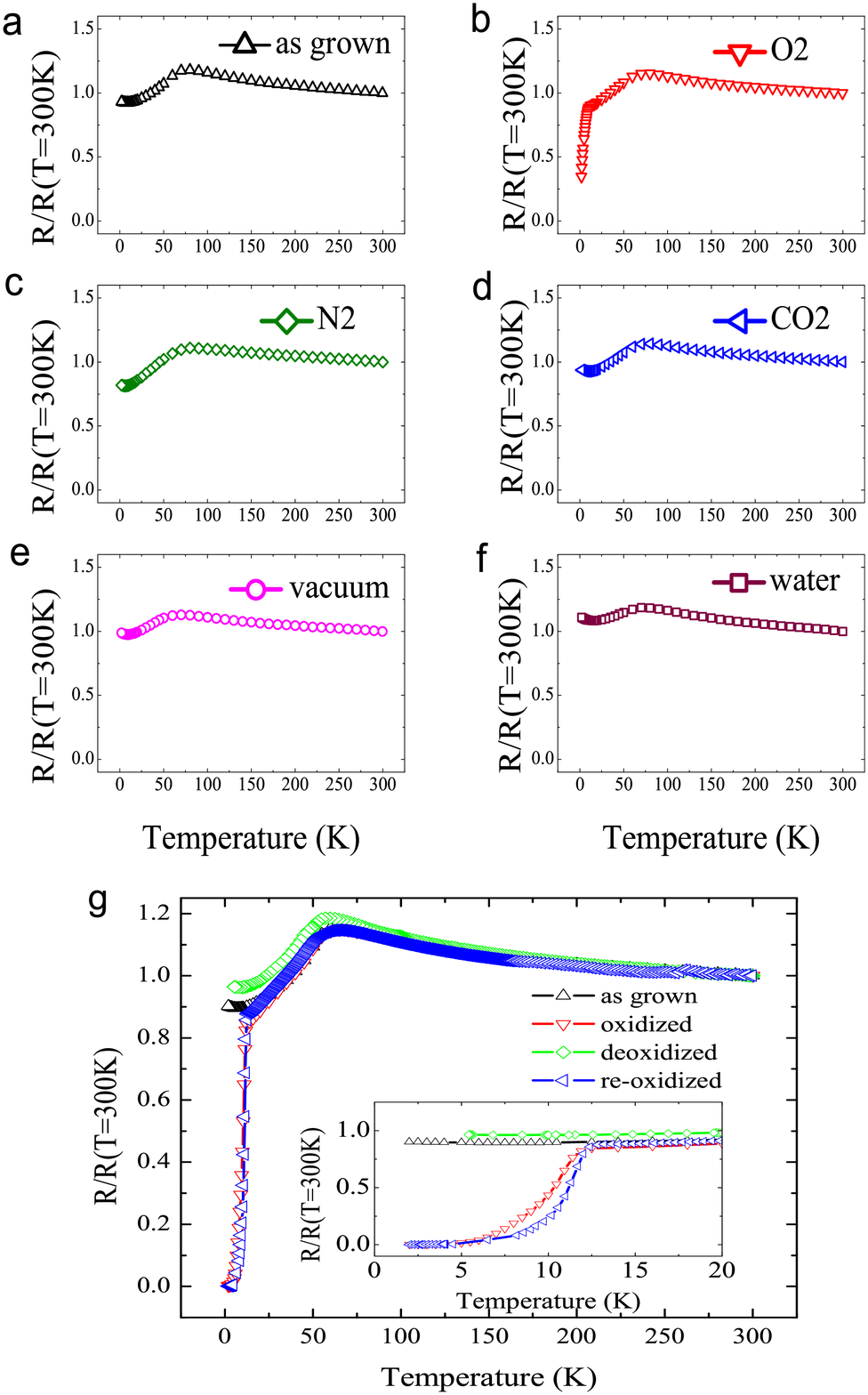}
\renewcommand{\figurename}{Figure}
\caption{\label{fig:fig3.eps} {\bf Effects of exposure to various
air components.} Separate FeTe samples were annealed in 100 $^0$C,
100 mTorr {\bf (b)} O$_{2}$, {\bf (c)} N$_{2}$, {\bf (d)} CO$_{2}$,
{\bf (e)} vacuum and exposed to {\bf (f)} water. Only the FeTe
sample annealed in O$_2$ shows superconducting transition near 10 K.
The other films show no big change from the as-grown FeTe films, see
{\bf(a)}. {\bf (g)} The same FeTe film sample was annealed in the
following sequence: (1) as-grown; (2) annealed in 100 $^0$C, 100
mTorr O$_2$ for 30 min; (3) annealed in 380 $^0$C, 10$^{-8}$ Torr
vacuum for 30 min. (4) re-annealed in 100 $^0$C, 100 mTorr O$_2$ for
30 min. Inset: a close up of low temperature region. It shows that
the superconductivity can be suppressed by annealing in vacuum and
restored by a follow up annealing in oxygen. }
\end{center}
\end{figure}


In order to determine what aspect of the exposure in air caused the
appearance of superconductivity, we conducted careful low
temperature annealing experiments. Separate FeTe samples were
annealed at 100 $^0$C in O$_{2}$, N$_{2}$, CO$_{2}$, vacuum and
exposed to water. For comparison, the resistivity measurement of the
as-grown FeTe films is shown in Fig. 3(a). The FeTe films annealed
in pure O$_{2}$ show a sharp drop of resistivity at around 10 K
similar to the air exposed sample indicating the onset of the
superconducting transition, see Fig. 3(b). FeTe films annealed in
N$_{2}$, CO$_{2}$ or vacuum have no sign of superconductivity down
to 2 K, see Fig. 3(c), 3(d), and 3(e). Water exposed films show a
substantial increase in resistivity but no sign of
superconductivity, see Fig. 3(f). Repeated anneals improved the
superconducting properties, most notably the sharpness and
completeness of the transition. However, adding subsequent anneals
of 50 minutes each increased both the room temperature resistivity
and degraded the superconducting transition. This degradation may
indicate the possibility of overdoping or may indicate some
degradation of the structure. Further, the oxygen doping is
reversible. Fig. 3(g) shows the resistivity of an FeTe film that was
first annealed in O$_{2}$ to be superconducting. Subsequently, the
sample was annealed in vacuum to drive out the excess oxygen and
this film was not superconducting with a resistivity vs. temperature
approximately the same as as-grown samples. Finally, a further
oxygen anneal restored the superconducting state. Taken together,
this set of annealing experiments provides strong evidence that the
fundamental change that leads to superconductivity in the FeTe films
is the reversible incorporation of oxygen.

We pursued X-ray absorption experiments at the Fe L edge and O K
edge to understand the role of the oxygen that is incorporated into
the FeTe structure. The TEY and TFY results at the Fe L edges were
identical (not shown), while the O K edge TEY looks similar to bulk
FeOx (not shown) but the TFY is dominated by the MgO substrate
demonstrating that TFY is sampling the entire bulk of the film. In
Fig. 4(a) we display the results for samples at three different
stages of oxidation under ambient conditions: after 4, 7 and 24
days. FeTe is expected to be charge balanced such that Fe is in a
nominal 2+ valence state. Consistent with this expectation, the
total florescence yield (TFY) of FeTe sample measured 4 days after
growth has an Fe L edge spectra which is closest in shape to that of
FeO, compared to any of the other bulk
FeO$_x$'s~\cite{Park01,Regan01,Kuepper01}, see the
inset~\cite{Park01}. In contrast, for the sample with the longest
exposure to air (24 days), the TFY Fe L edge spectra look most like
the Fe$_2$O$_3$. In particular, the spectra look most like
$\gamma$-Fe2O3 rather than $\alpha$-Fe$_2$O$_3$, but the two are
quite similar compared to the changes we observe with doping FeTe.
The FeTe film with intermediate exposure to air (7 days) has an Fe L
edge that appears to be a sum of roughly equal contributions from an
FeO like spectrum and an Fe$_2$O$_3$ like spectrum. The different
crystalline environment for Fe in FeTe, compared to the oxides,
rules out a reliable quantitative analysis of Fe$^{2+}$ and
Fe$^{3+}$ contributions. However, the qualitative result is clear,
that ambient oxidation drives the initial 2+ Fe ions to a 3+ like
nominal valence state. Our observation of the Fe valence states is
different from the literature on iron pnictides. In iron pnitides,
the Fe L edge spectra were reported to be metallic in nature; an
indication of weak electronic correlations~\cite{Yang01}. Our data
are thus consistent with claims that the iron chalcogenides are more
strongly correlated than the pnictides~\cite{Turner01,Tamai01} and
are also consistent with a larger moment in the
chalcogenides~\cite{Li01,Cruz01}.

\begin{figure}
\begin{center}
\includegraphics[width=78mm]{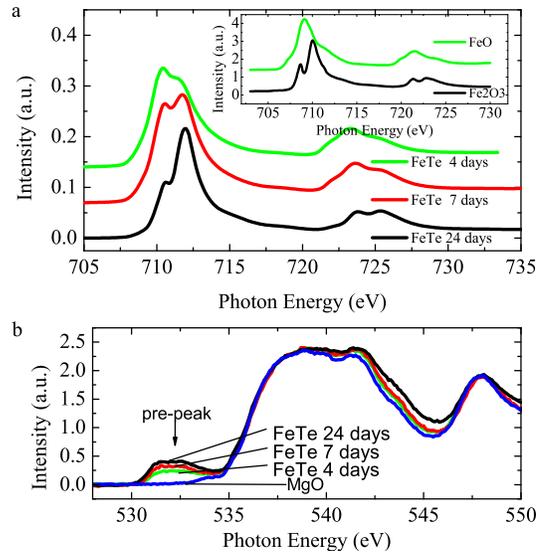}
\renewcommand{\figurename}{Figure}
\caption{\label{fig:fig5.eps} {\bf X-ray absorption spectroscopy of
FeTe films.} {\bf (a)}. The Fe L-edges show that the nominal valence
state of Fe increases with longer exposure to air, from 2+ to 3+, by
comparing with the FeO and Fe$_2$O$_3$ curves in the inset. {\bf
(b)}. The O K-edge spectra show a pre-peak which is a direct result
of Fe 3d orbitals hybridizing with O 2p orbitals, where the increase
in pre-peak intensity is proportional to the exposure time to air. }
\end{center}
\end{figure}

In Fig. 4(b), the TFY O K edge spectra provide further evidence that
oxygen is incorporated into the FeTe matrix. The bulk nature of the
TFY measurement causes the spectra to be dominated by the MgO
substrate which can be seen by comparison to the spectra from a bare
MgO substrate. However, the pre-peak in the energy range between 530
and 535 eV only appears with the FeTe films. Such a peak is a direct
result of Fe 3d orbitals hybridizing with oxygen 2 p
orbitals~\cite{deGroot01}. The relative photon energy position of
the FeO$_x$ pre-peak to the main MgO peak is in very good agreement
with the peak spacing reported in XAS studies of Fe$_{1-x}$Mg$_x$
O$_y$~\cite{Liu01}.


Does the valence state of Fe ions play an important role in creating
superconductivity in the Fe-chalcogenide system? Previous studies
seem to indicate that it does not. FeTe is a parent compound that
while poorly conducting, is antiferromagnetic and not
superconducting. In previous work, creating a superconductor from
FeTe involved substituting isovalent Se or S for some fraction of
the Te ions~\cite{Yeh01,Fang01}. This does not seem to involve
changing the valence of Fe or the charge balance of the active Fe
planes. However, the process for creating a superconductor we report
here is quite different - turning FeTe films into superconductors
involves incorporating oxygen into the structure interstitially and
not substitutionally. That observation combined with the X-ray
absorption data indicate that the nominal Fe valency changes as the
film becomes superconducting. Apparently either route will create a
superconductor out of the FeTe parent material. This is a
fundamental difference from the physics of the copper oxide high
temperature superconductors where hole concentration is universally
linked to the presence of superconductivity. It is, however,
consistent with the full range of data on the Fe-based
superconductors. The original Fe-based superconductor involved
doping the parent LaOFeAs compound by substitutionally doping F onto
the O site, a form of charge doping~\cite{Hosono01}. However, other
Fe-based superconductors can be created with stoichiometry that does
not include any apparent charge doping but presumably a lattice
distortion from the parent compound~\cite{Yeh01,Hsu01,Wang01}. Thus
the overall picture that one must move away from a specific point in
the phase diagram rather than moving toward a particular point
receives strong confirmation in the FeTe system.

Experimentally, the ability to reversibly turn on superconductivity
with suitable, low temperature annealing steps opens the door for a
variety of interesting, well controlled experiments. This could be
particularly important for surface related probes such as
photoemission to determine key electronic aspects of the
superconducting state.

Our results are also important in light of a finding recently
published by Han et al. who found superconductivity in films of FeTe
and attributed the result to in-plane, tensile strain.~\cite{HanY01}
In our case, the annealing experiments make a clear case that
superconductivity is induced by incorporation of oxygen. An
examination of this reference does not seem to rule out a similar
cause for superconductivity in their samples. Oxygen in those films
was never measured and they were grown with a background pressure an
order of magnitude higher than in our work. Our films covering the
same range of thickness were never superconducting immediately out
of the growth chamber. Of course direct measurements would be
necessary to know and it could be possible that either mechanism
could produce superconducting versions of FeTe.


In summary, by introducing excess oxygen at low temperature,
superconductivity was induced in the FeTe films with an onset
temperature near 10 K. The oxygen doping process is reversible.
X-ray absorption spectroscopy indicates that the FeTe films become
superconducting concurrent with a nominal Fe valence state increase
from 2+ to 3+. This method for creating a superconductor is
drastically different from the better known procedure of
substituting isovalent Se or S for Te. Thus superconductivity occurs
in at least two very different places in the phase space of strain
and Fe valence. A proper theory for superconductivity in these
materials must account for this new conceptual phase diagram.


{\bf Acknowledgements}

We thank F. Rueckert for assistance in collecting the synchrotron
diffraction data. We would like to thank E. Negusse for assistance
during absorption measurements at the U4B beam line. This work is
supported by the US-DOE through contract \# DE-FG02-00ER45801. Use
of the National Synchrotron Light Source, Brookhaven National
Laboratory was supported by the Office of Science, Office of Basic
Energy Sciences, U.S. Department of Energy under Contract No.
DE-AC02-98CH10886.


\end{document}